\documentclass[12pt]{iopart}
\usepackage{graphicx}
\usepackage{dcolumn}
\usepackage{bm}

\def\bbbc{{\mathchoice {\setbox0=\hbox{$\displaystyle\rm C$}\hbox{\hbox
to0pt{\kern0.4\wd0\vrule height0.9\ht0\hss}\box0}}
{\setbox0=\hbox{$\textstyle\rm C$}\hbox{\hbox
to0pt{\kern0.4\wd0\vrule height0.9\ht0\hss}\box0}}
{\setbox0=\hbox{$\scriptstyle\rm C$}\hbox{\hbox
to0pt{\kern0.4\wd0\vrule height0.9\ht0\hss}\box0}}
{\setbox0=\hbox{$\scriptscriptstyle\rm C$}\hbox{\hbox
to0pt{\kern0.4\wd0\vrule height0.9\ht0\hss}\box0}}}}

\newcommand{\beq}{\begin{eqnarray}}
\newcommand{\eeq}{\end{eqnarray}}

\newcommand{\bQ}{{\bf Q}}
\newcommand{\bk}{{\bf k}}

\newcommand{\bq}{{\bf q}}

\newcommand{\beqa}{\begin{eqnarray}}
\newcommand{\eeqa}{\end{eqnarray}}

\renewcommand{\Re}{{\rm Re}}

\def\etal{{\sl et al.}}

\def\nonum{\nonumber \\}

\def\nonum{ \nonumber \\}

\def\urusi{URu$_2$Si$_2$ }

\begin{document}

\title{Charge density wave in hidden order state of URu$_2$Si$_2$}

\author{Jung-Jung Su$^{1,2}$, Yonatan Dubi$^1$, Peter W\"olfle$^3$ and Alexander V. Balatsky$^{1,2}$}


\address{$^1$Theoretical Division, Los Alamos National Laboratory, Los Alamos, New Mexico 87545, USA}
\address{$^2$Center for Integrated Nanotechnologies, Los Alamos National Laboratory, Los Alamos, New Mexico 87545, USA}
\address{$^3$ Institute for Theory of Condensed Matter and Center for Functional Nanostructures, Karlsruhe Institute of Technology, D-76128 Karlsruhe, Germany}

\ead{jungksu@lanl.gov}

\begin{abstract}
We argue that the  hidden order state in \urusi  will  induce a charge density wave. The modulation vector of the charge density wave will be twice that of the hidden order state, $Q_{CDW} = 2Q_{HO}$. To illustrate how the charge density wave arises we use a Ginzburg-Landau theory that contains a coupling of  the charge density wave amplitude to the square of the HO order parameter $\Delta_{HO}$. This simple analysis allows us to predict the  intensity and temperature dependence of the charge density wave order parameter in terms of the susceptibilities and coupling constants used in the Ginzburg-Landau analysis.

\end{abstract}

\maketitle

\section{Introduction}
The exact nature of the  Hidden Order (HO) in \urusi has been a well documented puzzle for more then 20 years \cite{Palstra,Schlabitz,Maple}. The onset of HO is seen in the specific heat as a sharp  mean field like transition at $T_{HO} = 17.5K$.  However, a systematic search  of any magnetic or structural transition at $T_{HO}$  yielded negative results, thus suggesting the term 'Hidden Order' to highlight the missing connection between textbook mean field features, e.g. in the specific heat and the lack of any observable order parameter.  Various proposals made to account for this phase generally fall within two categories that can be broadly summed up as (i) localized intra unit cell ordering and (ii) extended momentum space ordering. The first category, in which speculations were made to the effect that the HO is characterized by some local  ordering \cite{Chandra,Santini,Kasuya,Sikkema,Cricchio,Haule}, was prompted by the large specific heat change and large entropy release at the HO transition,  the latter on the scale of $\delta S \approx 0.2 R \ln 2$.  Alternatively, the HO transition was viewed as an ordering of itinerant degrees of freedom, thus intrinsically being a  momentum space phenomenon with the ordering occurring in momentum space \cite{Ikeda,Varma,Balatsky,Elgazzar,Dubi}.

 We still do not know which of these models are closer to the true nature of the HO state. New hints emerging from momentum resolved spectroscopies point to the existence of a momentum space instability as an important ingredient, if not a key ingredient, in the HO puzzle. First, the new as well as old neutron scattering data \cite{Broholm,Wiebe,Ikeda2} provide tantalizing hints that indeed the spin dynamics drastically changes in the HO state. New sharp resonant features appear to develop in the HO state in the spin susceptibility as seen by inelastic neutron scattering, at both commensurate momenta ${\bf Q} = (1,0) , (0,1) \pi/a$  \cite{Bourdarot} at $\omega  = 2$ meV  and at  incommensurate momenta ${\bf Q^*} = (0.6,0), (0,0.6) \pi/a$ at $\omega_r = 5$ meV \cite{Bourdarot,Wiebe,Balatsky}, indicating changes in the spin dynamics below $T_{HO}$. Second, recent Scanning Tunneling Microscopy (STM) results revealed the onset of a gap like feature in the tunneling spectra, and Quasiparticle Interference (QPI) data reveal the onset of a hybridization feature in the quasiparticle band structure of \urusi \cite{Schmidt} that develops near or at the incommensurate vector $Q^*$.  The hybridization features seen in the tunneling spectra,  in  quasiparticle dispersion in STM and in neutron scattering resonance, all have the same energy  scale of  $\sim5 $ meV at characteristic momentum transfers $Q^*$. Therefore, it is at least a plausible suggestion that the HO is a phenomenon that is controlled by momentum space instability. Such a momentum space instability at $Q^*$ would produce a  modulation of the order parameter $\Delta_{HO}({\bf x}) =\Delta_{HO} \exp(i {\bf Q^*} {\bf x})$.

 A question that naturally arises from these observations is: if the HO is an incommensurate (or a commensurate) wave phenomenon of some sort and would this wave produce a charge density wave (CDW) ? In this communication we prove that a spatially modulated HO parameter will indeed lead to a CDW in the HO state. To prove this point we employ a Ginzburg-Landau (GL) analysis including a coupling between the {\em local } charge density modulation $\rho(x)$  and the {\em square} of the HO order parameter $\Delta_{HO}$. We use the fact that for any order parameter its  square amplitude will be a simple scalar and hence in the free energy analysis it can couple to the charge density via
 \beqa
 F_{int} =- \int d^d{\bf x} \lambda \ \rho({\bf x}) \, \Delta_{HO}({\bf x}) \Delta_{HO}^*({\bf x})
 \label{EQ:coupling1}
 \eeqa
 with coupling strength $\lambda$ that can be of either sign. $\rho({\bf x})$ and $\Delta_{HO} ({\bf x})$ are the charge density and the hidden order parameters at position $x$. The charge stiffness free energy is given by the form:
 \beqa
 F_{CDW} = \sum_{\bq} f_q \, | \rho_{\bq}|^2.
 \label{EQ:charge1}
 \eeqa
where $f_q=a_1/q^2 + a_2 \,q^2 + a_3 >0 $ for all momenta\cite{Turgut}. The $a_1$ term describes the Coulomb repulsion, the $a_2$ term describes the gradient energy costs for density modulations and the $a_3$ term describes a short range term. The positivity of $f_q$ insures that there is no CDW in the absence of the coupling. From this point on we take $f_q=\chi/2>0$ for simplicity. As we show below, the amplitude of the induced CDW modulation is proportional to $\lambda$ and inversely proportional to the charge stiffness $\chi$. Minimization of the total free energy  $F=F_{CDW}+ F_{HO}+F_{int}$ leads to
 \beqa
 \rho_{2 \bq_1} = \frac{\lambda}{\chi} \Delta_{HO}^2 (\bq_1)~~,
 \label{EQ:charge2}
 \eeqa
which implies that the modulation wave vector of the CDW is {\em double that of the HO modulation} in real space. This observation is the main result of the paper.
If the HO modulation is commensurate with the lattice then the CDW will also be commensurate (case of $Q$ modulation). For instance, if the HO state is a single unit cell order phenomenon without translational symmetry breaking, the CDW momenta are identical with the main Bragg peaks of the lattice. On the other hand if the HO state is incommensurate, so will be the CDW (case of $Q^*$ modulation).  For the HO modulated at an incommensurate wavevector like ${\bf Q^*} = (0.6,0), (0,0.6) \pi/a$ the CDW modulation will be at $\bQ_{CDW} = (1.2,0), (0,1.2) \pi/a$. Another immediate consequence of the GL analysis is that  the intensity of the CDW will scale as $\rho_{2 \bq_1} \sim \Delta_{HO}^2 (\bq_1) \sim |T-T_{HO}|$, since the HO order parameter sets in as at a well defined mean field transition.


\begin{figure}[t]
\begin{center}
\includegraphics[width=1.1\linewidth,angle=0,keepaspectratio]{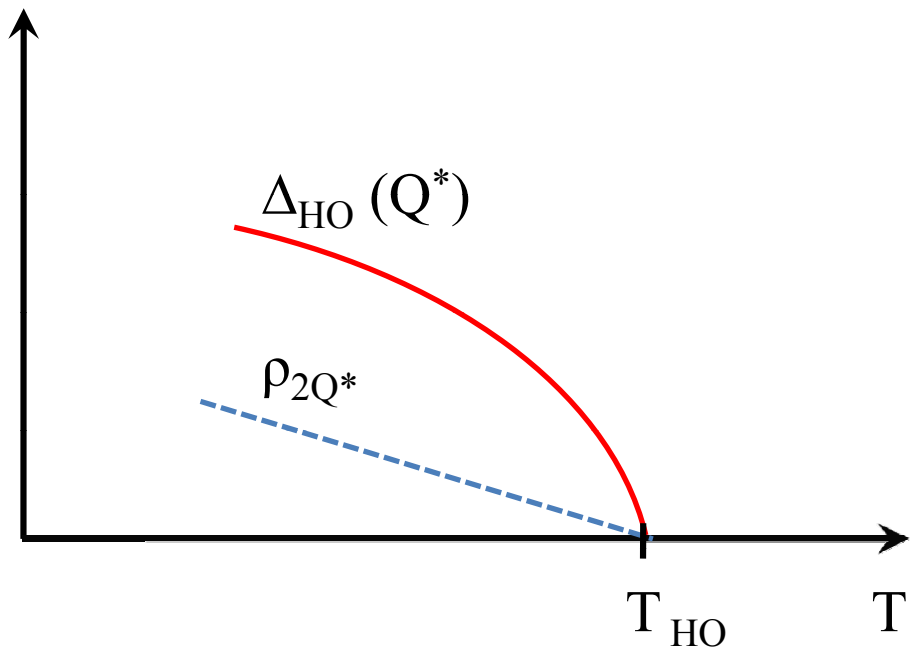}
\end{center}

\vskip-.7cm
\caption{(Color online)
 Illustration of the onset of CDW amplitude $\rho_{2Q*} \sim |T-T_{HO}|$ and HO parameter $\Delta_{HO} \sim |T-T_{HO}|^{1/2}$ below HO transition. The CDW order is induced by onset of HO and opens up with higher power of $T-T_{HO}|$
}\label{fig:CDW}
\end{figure}

\section{Ginzburg-Landau Model for CDW}

Recently two of us proposed that the HO is a hybridization wave with a modulation of $Q^*$ \cite{Dubi}. In our model the HO in \urusi originates from an indirect excitonic condensation of $d-$band holes bound to local $f-$band electrons. In general, such hybridization order takes the form
\begin{eqnarray}
\Delta_{HO} ({\bf k}) = \sum_{\bf q} V_{{\bf q+k},{\bf q}} \ \langle \, c^{\dagger}_{{\bf q}} f_{{\bf q+k}} \rangle
\end{eqnarray}
In particular, we  argue that the coupling is mainly between electrons (and holes) with momenta ${\bf Q^*}/2$ (and ${\bf -Q^*}/2$) and vice-versa, where ${\bf Q^*}/2=(0.3,0) \pi/a$ . This coupling induces a modulation wave vector  of ${\bf Q^*}$ in the hidden order state. In total there are four incommensurate wave vectors ${\bf Q_{i}}$, ${\bf Q_{1}}={\bf Q^*}$ and the three equivalent vectors in tetragonal symmetry, rotated in the (a,b)-plane by multiples of $\pi /2$ . We will assume in the following that the HO phase forms a monodomain characterized by a single wave vector direction such that ${\bf Q_{i}}=\pm {\bf Q^*}$ . The corresponding Ginzburg-Landau (GL) energy functional for the HO is \cite{Turgut}
\beqa
F_{HO} &=&
\sum_{\bf k} c_{{\bf k}} |\Delta_{HO} ({\bf k})|^2 \nonumber \\
&& + \frac{1}{2} \alpha \, |\Delta_{HO} ({\bf k})|^2
+ \frac{1}{4} \beta \, |\Delta_{HO} ({\bf k})|^4
\eeqa
where the first term is the kinetic energy and $c_{{\bf k}}=c \sum_{\bf Q_{i}} ({\bf k-\bf Q_{i}})^2$ is the stiffness of the hidden order field, the coefficient of the second order term changes sign at $T=T_{HO}$, $\alpha=a(T-T_{HO})$, and $\beta$ is a positive constant. It is clear that without the coupling with other degrees of freedom, $F_{HO}$ favors of a modulation with momentum $\pm Q^*$, noting that $T_{HO}$ is largest for these momentum values. The free energy associated with a charge density modulation is given by
\beqa
F_{CDW} = \sum_{\bf q} \frac{1}{2} \, \chi \, |\rho_{\bf q}|^2.
\eeqa
The HO and the charge degrees of freedom are coupled through
\beqa
F_{int} = -\lambda \, \sum_{k,q} \rho_{\bf q} \Delta_{HO}({\bf k-q}) \Delta^*_{HO} ({\bf k}) + c.c.~, \label{F_int}
\eeqa
which is simply the Fourier transform of Eq.~\ref{EQ:coupling1}. The total energy function is
\beqa
F = F_{HO}+F_{CDW}+F_{int}~~.
\eeqa
Minimizing the energy functional yields the equation set
\beqa
0  &=& \frac{\partial F}{\partial \Delta_{\bk}^{HO}}
= -\lambda \sum_{\bq} \rho_{\bq} \left( \Delta^*_{HO}({\bk}+{\bq}) + \Delta^*_{HO}({\bk}-{\bq}) \right) \nonumber \\
&& \hspace{1.5cm} +2 C \left({\bk} -{\bQ}^{*} \right)^2 \Delta_{HO}^* ({\bk}) \nonumber \\
&& \hspace{1.5cm} + \alpha \, \Delta_{HO}^* ({\bk})
+ \beta \, |\Delta_{HO} ({\bk})|^2 \Delta_{HO}^* ({\bk}) \label{Deltaeq}\\
0 &=& \frac{\partial F}{\partial \rho_{\bq}}
= \chi \rho_{\bq}
-\lambda \sum_{\bk} \, {\Re}\left[\Delta_{HO} ({\bk}+{\bq}) \Delta^*_{HO}({\bk}) \right]
\label{rhoqeq}
\eeqa

Near the critical temperature $T_{HO}$, the only components of the HO parameter are $\Delta_{HO}(+{\bf Q^*})$ and $\Delta_{HO}({\bf -Q^*})$, yielding possible momentum values for ${\bf q}$ in Eq.~
\ref{F_int}, ${\bf q}={\pm \bf 2Q^*}$, and therefore a single Fourier component for the charge modulation $\rho_{2Q^*}$. More precisely, the only vanishing contribution to $\Delta_{HO}(q)$ and $\rho_q$ are:
\beqa |\Delta_{HO}( \pm {\bf Q^*})|
&=&\left( - \frac{\alpha}{\beta-\lambda^2/\chi} \right)^{1/2} \nonum
|\rho_{2 {\bf Q^*}}|
&=&\left| -\frac{\alpha\lambda}{\beta \chi-\lambda^2}\right|=\frac{\lambda}{\chi}|\Delta_{HO}( {\bf Q^*})|^2 ~~.\label{MF solutions} \eeqa
The amplitude of $\Delta_{HO}(\pm {\bQ}^*)$ increases with respect to the $\lambda=0,~\rho=0$ solution, resulting in a decrease of the free energy (for negative $\lambda$), indicating that a CDW is formed. The amplitude of the resulting CDW is \emph{linear} in $T_c-T$, which is a direct experimental prediction.

\begin{figure}[t]
\begin{center}
\includegraphics[width=1.1\linewidth,angle=0,keepaspectratio]{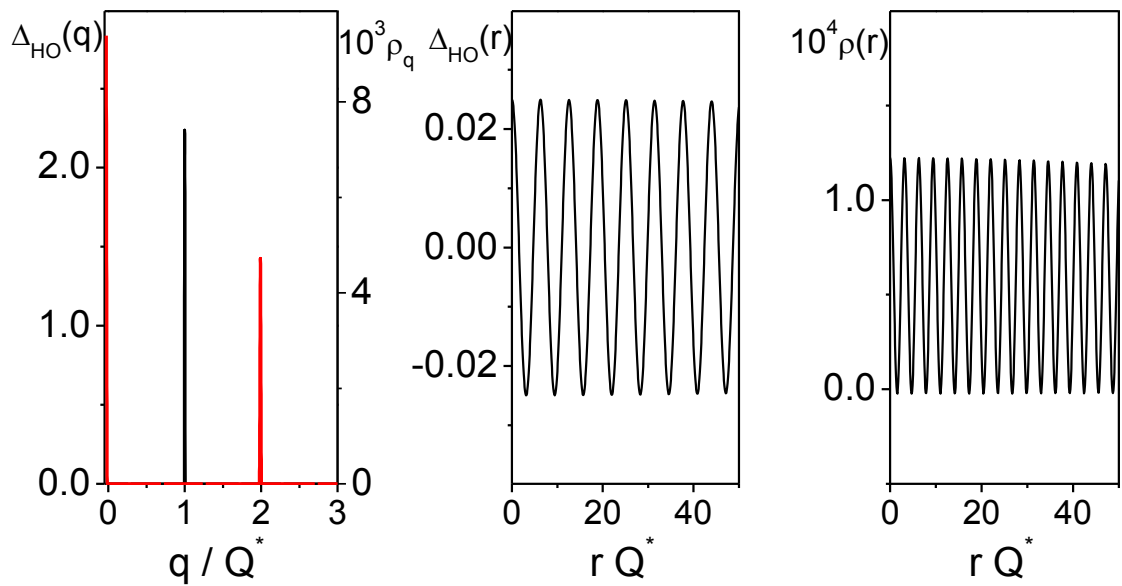}
\end{center}
\vskip-.7cm
\caption{(Color online)
Numerical solution for hidden order parameter and charge density wave. The parameters used here are $c=50.0, \alpha=-0.005, \beta=0.001, \chi=1.0, \lambda=0.05$. The left panel shows the $\Delta_{HO} (q)$ (black) and $\rho_q$ (red). The modulation of $\Delta_{HO} (q)$ at $Q^*$ induces finite charge at $2Q^*$. The middle and right panels are the hidden order and the charge modulation in real space.
}\label{fig:C=50}
\end{figure}

\begin{figure}[t]
\begin{center}
\includegraphics[width=1.1\linewidth,angle=0,keepaspectratio]{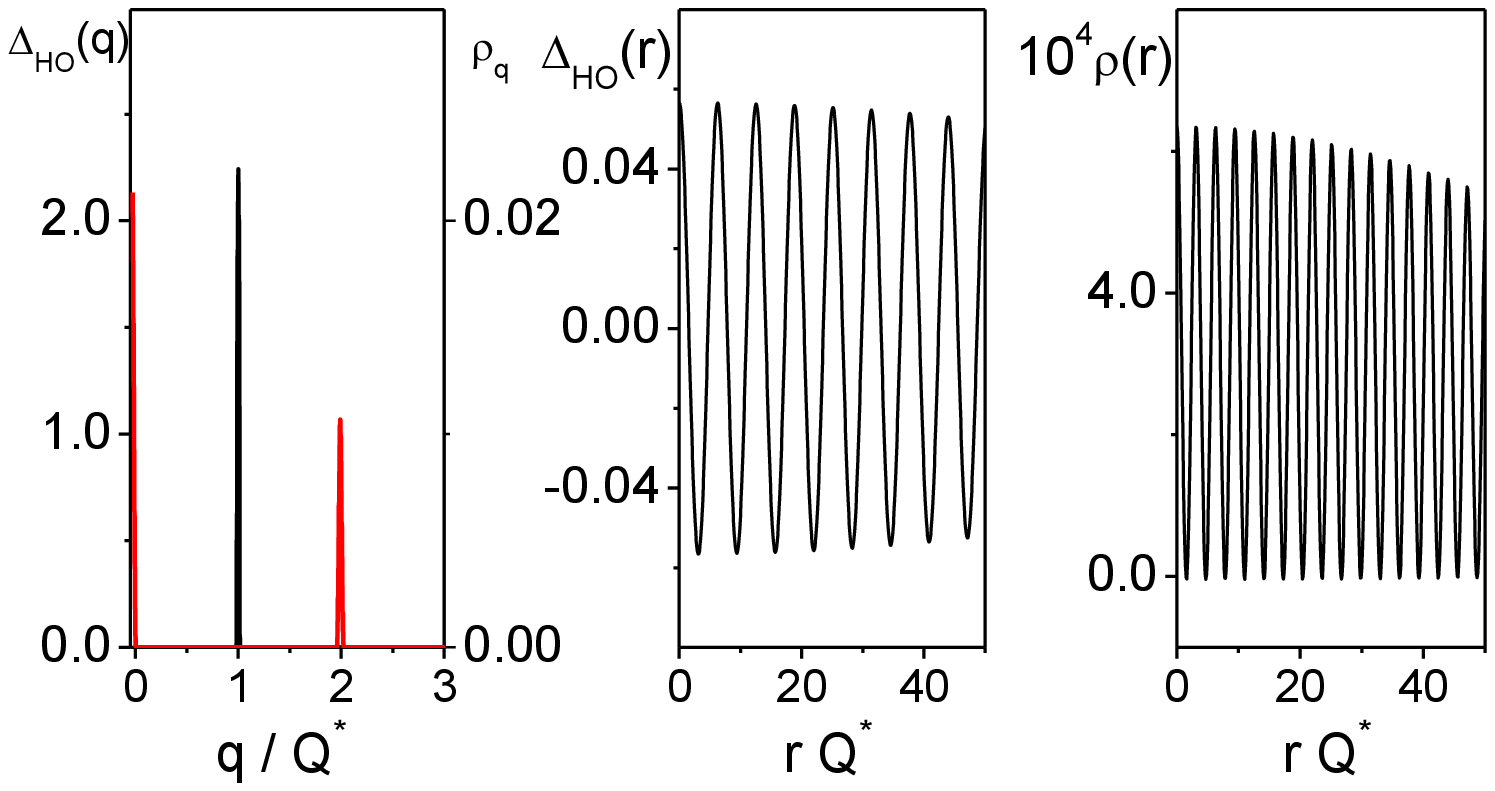}
\end{center}
\vskip-.7cm
\caption{(Color online)
The same as Fig.\ref{fig:C=50} except for $c=10.0$.
}\label{fig:C=10}
\end{figure}

\begin{figure}[t]
\begin{center}
\includegraphics[width=1.1\linewidth,angle=0,keepaspectratio]{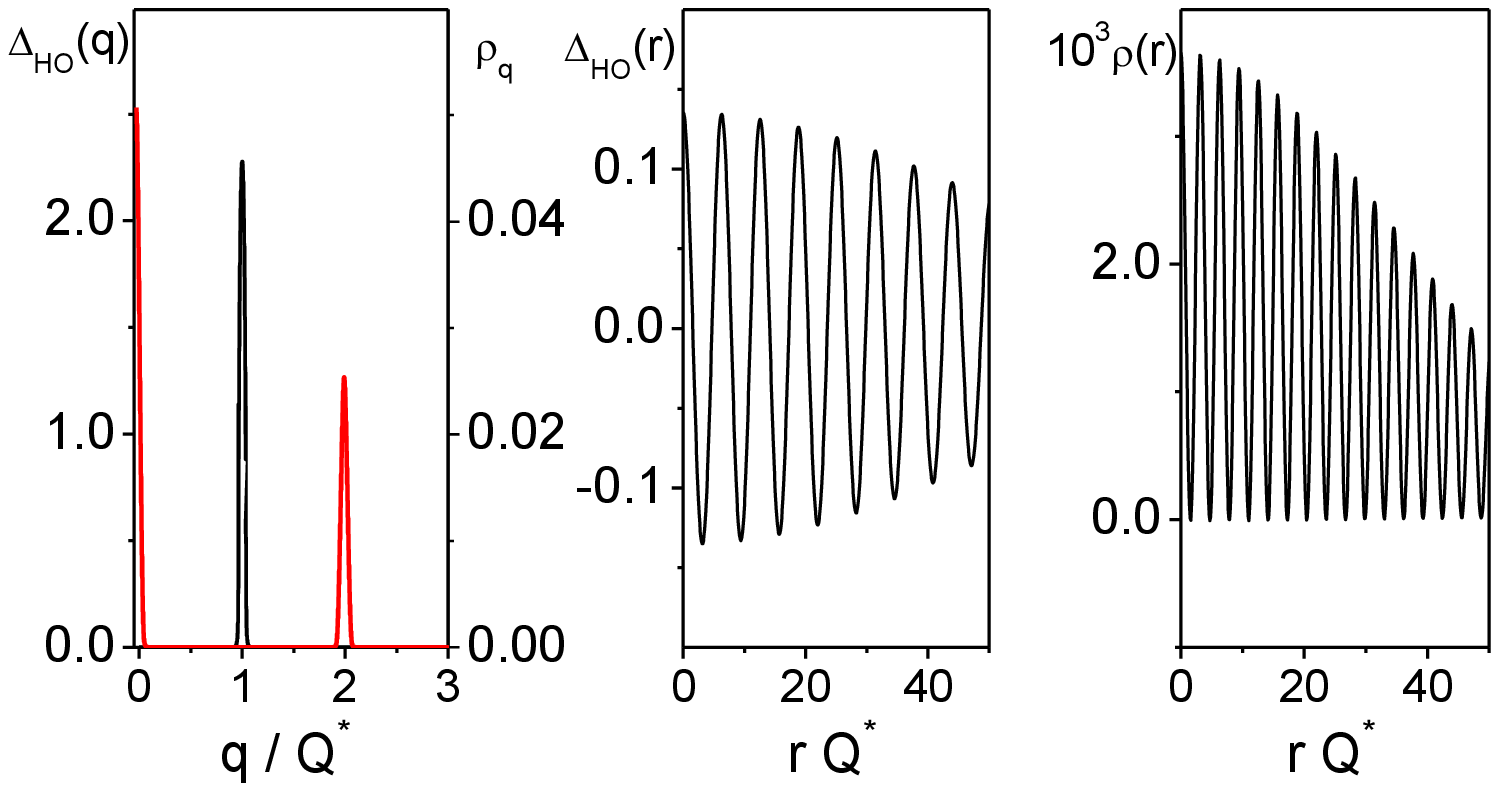}
\end{center}
\vskip-.7cm
\caption{(Color online)
The same as Fig.\ref{fig:C=50} except for $c=2.0$.
}\label{fig:C=2}
\end{figure}

In Fig. \ref{fig:C=50}~,\ref{fig:C=10} and \ref{fig:C=2} we plot the numerical solution for Eq.~(\ref{Deltaeq}-\ref{rhoqeq}) for the parameters $\alpha=-0.005, \beta=0.001, \chi=1.0, \lambda=-0.05$ and $c=50,10,2$ respectively (which effectively correspond to the weak, intermediate and strong coupling regimes).
The numerical calculations are consistent with the results obtained above. In the weak coupling regime, we obtain a CDW oscillating with wave vectors around $2Q^*$ in all three cases. For large stiffness ( Fig.~\ref{fig:C=50} ), $\Delta_{HO}$ oscillates with a single wave vector $\bf Q^*$ and is not affected significantly by the coupling with the CDW. The CDW oscillates with a rapid oscillation $2Q^*$ on top of  a slow modulation. When the stiffness decreases ( Fig.~\ref{fig:C=10},\ref{fig:C=2} ), the peaks in $\Delta_{HO} (q)$ and  $\rho_q$ broaden up. Interferences between momenta within the peak width cause the modulation to vanish at large $r$.

\section{Conclusions}
In this paper we showed that in the presence of a HO parameter with modulation wavevector $Q^*$, a charge density wave with modulation wavevector $2Q^*$ emerges, provided an attractive coupling between the HO parameter amplitude (squared) and the charge density modulation exists in the free energy. We predict that the amplitude of the resulting CDW grows linearly in temperature below $T_{HO}$. We hope that our results will stimulate a detailed search for the proposed CDW. The natural techniques that would allow for CDW observatoin would be neutron scattering, x ray scattering and local probes. In princilple the optical conductivity and ARPES might also see effects of band changes due to CDW formation.   If CDW is observed it may shed further light on the nature of the hidden order state in \urusi.

\ack
The authors enjoyed fruitful discussions with J. C. Davis, M. J. Graf, B. Gaulin, G. Kotliar, G. Luke, E. Hassinger and J. Mydosh. This work was supported by the US Dept. of Energy at Los Alamos National Laboratory under contract No. DE-AC52-06NA25396 and by UCOP TR01 and by the DFG research unit "Quantum phase transitions" (PW).

\end{document}